\documentclass[12pt,letterpaper]{article}
\pdfoutput=1

\usepackage{amsfonts}
\usepackage{amssymb}
\usepackage{graphicx}
\usepackage{latexsym}
\usepackage{amsmath}
\usepackage{color}
\newcommand\openone{\leavevmode\hbox{\small1\normalsize\kern-.33em1}}
\def\ee{\end{eqnarray}}

\def\nn{\nonumber}

\newcommand{\<}{\langle}
\renewcommand{\>}{\rangle}

\newcommand{\be}{\begin{eqnarray}}
\newcommand{\en}{\end{eqnarray}}
\newcommand{\bea}[1]{\left(\begin{array}{#1}}
\newcommand{\ena}{\end{array}\right)}
\newcommand{\ba}{\begin{eqnarray}}
\newcommand{\ea}{\end{eqnarray}}
\newcommand{\CL}{\mathcal{L}}
\newcommand{\CM}{\mathcal{M}}
\newcommand{\CO}{\mathcal{O}}

\usepackage{hyperref}

\newcommand\lrpar{\raise .8ex\hbox{$^\leftrightarrow$} \hspace{-9pt}
\partial}
\newcommand\lpar{\raise .8ex\hbox{$^\leftarrow$} \hspace{-9pt}
\partial}
\newcommand\rpar{\raise .8ex\hbox{$^\rightarrow$} \hspace{-9pt}
\partial}

\begin{document} 

\vspace{5mm}
\vspace{0.5cm}
\begin{center}

\def\thefootnote{\fnsymbol{footnote}}

{\Large \bf Model Independent Direct Detection Analyses}
\\[0.5cm]
{\large  A.~Liam Fitzpatrick$^1$,  Wick Haxton${^2}$, Emanuel Katz$^{1,3,4}$, Nicholas Lubbers$^3$, Yiming Xu$^3$}


{\it $^1$ Stanford Institute for Theoretical Physics, Stanford University, Stanford, CA 94305} \\
{\it $^2$ Dept. of Physics, University of California, Berkeley, 94720,
and Lawrence Berkeley National Laboratory} \\
{\it $^3$ Physics Department, Boston University, Boston, MA 02215, USA} \\
{\it $^4$ SLAC National Accelerator Laboratory, 2575 Sand Hill, Menlo Park, CA 94025, USA.}

\end{center}

\vspace{.8cm}

\hrule \vspace{0.3cm}
{\small  \noindent \textbf{Abstract} \\[0.3cm]
\noindent


Following the construction of the general 	effective theory for dark matter direct detection in \cite{us}, we perform an analysis of the experimental constraints on the full parameter space of elastically scattering dark matter.  We review the prescription for calculating event rates in the general effective theory and discuss the sensitivity of various experiments to additional nuclear responses beyond the spin-independent (SI) and spin-dependent (SD) couplings: an angular-momentum-dependent (LD) and spin-and-angular-momentum-dependent (LSD) response, as well as a distinction between transverse and longitudinal spin-dependent responses.  We consider the effect of interference between different operators and in particular look at directions in parameter space where such cancellations lead to holes in the sensitivity of individual experiments.  We explore the complementarity of different experiments by looking at the improvement of bounds when experiments are combined.  Finally, our scan through parameter space shows that within the assumptions on models and on the experiments' sensitivity that we make, no elastically scattering dark matter explanation of DAMA is consistent with all other experiments at 90\%, though we find points in parameter space that are ruled out only by about a factor of 2 in the cross-section.

\vspace{0.5cm}  \hrule
\def\thefootnote{\arabic{footnote}}}
\setcounter{footnote}{0}

\section{Introduction}
\label{sec:intro}

The idea that the dark matter (DM) component of our universe is a new particle, or perhaps a new of set of particles, remains very attractive.  Current and future direct detection experiments 
offer the exciting possibility of obtaining the most complete picture of the properties of this particle.  Traditionally, the DM particle has been viewed mostly through the WIMP framework, 
whereby the DM particle is intimately connected with theories of electroweak breaking.  Thus, theories of DM have been written in terms of relativistic degrees of freedom relevant for
TeV scale physics.  On the other hand, the DM detection experiments measure the signal due to a non-relativistic collision of DM with nuclei in the body of the detector.  
When comparing experiments to each other, it is therefore more prudent to use a description that does not suffer from model-building biases coming from assumptions about
physics at a much higher energy scales.  Indeed, such assumptions have often lead to a view that there are only two possibilities for the interaction between nuclei and DM.  Namely,
that the coupling is either spin-independent (SI) or spin dependent (SD).  

In a previous paper, extending the work of \cite{fanEFT},we argued that with minor assumptions, there are, in fact, several non-relativistic operators
consistent with the rules of QM, Galilean-invariance, and CP symmetry, which describe possible nucleon-DM interactions.  An important difference with the standard paradigm is that we allowed 
for momentum dependent interactions as well (to second order in the momentum).  Such interactions are well-motivated, and fully relativistic models for these have been 
studied earlier as a way to reconcile the various experiments with each other.   For example, the DM particle could be composite (after all, most visible matter takes the form of atoms), and its interaction with nuclei
could proceed through dipole or charge-radius type operators sensitive to its size.  The advantage of this framework is that it is essentially model-independent and makes use only
of the ingredients that DM experiments are probing.  The operators we focus on also apply for DM particles of any spin.  Any high energy relativistic DM model 
can be 
translated into a particular linear combination of the EFT operators.  On the other hand, the non-relativistic operators form a general basis for the interactions most 
directly probed by the experiments themselves.  In many ways, these operators are analogous to the precision electroweak observables, which were useful in parameterizing 
the results of precision experiments measuring the properties of the W and Z in a model independent way.  Thus, instead of thinking of target nuclei in terms of SI and SD features alone,
a more model independent analysis should be one which considers the nuclear responses coming from the non-relativistic interactions.

Interestingly, we found that some of the new momentum-dependent interactions give rise to novel nuclear responses not considered in the DM literature.  
In particular, there are five different response functions which enter.
They are:  the standard SI response, two types of spin-dependent responses, SD1 and SD2 (a particular combination of which is the usual SD), a response which
for certain elements is significantly dependent on the angular-momentum of unpaired nucleons within the nucleus (LD), and a novel response which depends on a product
of spin and angular momentum (LSD).  In addition, interference is possible between operators with similar quantum numbers.  In certain regions of parameter space,
this interference could significantly alter the strength with which DM couples to a particular target nucleus.  

In this paper we will first present the non-relativistic operators and the five nuclear responses they elicit.  Our goal will then be to explore the parameter space of these operators,
given current experimental bounds.   In particular, as various targets will couple differently to DM, our focus will be to determine which experiments need to be combined in
order to optimize sensitivity to a given DM scenario.  It is important to check whether there are potential ``blind spots"  with the targets currently being used,  In addition,
it is useful to know which targets compliment each other the most, yielding better sensitivity in the future.  

Our paper is organized as follows.  In Sec. \ref{sec:review} we review the non-relativistic effective theory, listing the operators and present our framework for analysis of
the responses.  In Sec. \ref{sec:constraints} we present the current experimental constraints on the parameter space, emphasizing where there are potential gaps
in sensitivity.  We also provide a qualitative description of the various responses.  Finally, in Sec. \ref{sec:discuss} we summarize our results and conclude.

\section{Review of Effective Theory}
\label{sec:review}

\subsection{Building Blocks of the Theory}

Interactions in the effective theory are constructed from a small number of ``building blocks'' that are directly connected to the non-relativistic process of a dark matter particle scattering off
of a nucleon inside a nucleus. 
The first of these building blocks are the momenta of the particles in the collision.  Since we are interested in two-to-two scattering, there are a priori four different momentum
vectors, namely $p, k$ for the incoming dark matter and nucleon momenta, respectively, and $p',k'$ for their outgoing momenta.  However, the constraints of momentum-conservation
and independence of inertial frames allows us to reduce these from four to two independent momenta.  Without loss of generality, one may take these
to be $\vec{q}=\vec{p}~'-\vec{p}$, which is the momentum transfer of the collision, and $\vec{v} = \vec{v}_{\chi, \rm in}- \vec{v}_{N, \rm in}$, which is the incoming DM velocity in the
nucleon rest frame.  While these are both manifestly frame independent quantities, they are not the most convenient combinations to consider, because the most general interaction
built out of them is not Hermitian.  Since Hermitian conjugation interchanges in and out states, one effectively has $\vec{q} \rightarrow - \vec{q}$ and $\vec{v} \rightarrow  \vec{v}_{\chi, \rm out} - \vec{v}_{N, \rm out} = \vec{v} + \frac{\vec{q}}{\mu_N}$, where $\mu_N$ is the dark-matter-nucleon reduced mass. Consequently, it is natural to work instead with the quantities
\be
i \vec{q} &\textrm{and} & \vec{v}^\perp = \vec{v} + \frac{\vec{q}}{2 \mu_N} .
\ee
In addition, the spins $\vec{S}_\chi$ and $\vec{S}_N$ of the dark matter and the nucleons, respectively, are Hermitian operators that can appear in interactions. For spin-$\frac{1}{2}$ fermions (i.e. for the dark matter in many models, and the nucleons in any case), these are just Pauli sigma matrices.  The most general set of interactions for elastically scattering dark matter are the rotational invariants that can be constructed out of these four basic building blocks:
\be
 \vec{S}_\chi, \vec{S}_N, i \vec{q}, \textrm{  and  } \vec{v}^\perp.
 \label{eq:blocks}
 \ee

\subsection{List of Effective Operators}
\label{sec:EFTlist}

In practice, we will not study all possible rotationally-invariance combinations of the building blocks in eq.~(\ref{eq:blocks}), but instead will make the further restriction that they can arise from the exchange of a spin-0 or spin-1 field.  Effectively, this means including at most two powers of the spins and/or velocity.  This leaves us with the following $T$-even operators, classified according to symmetry into different groups such that operators in different groups will not interfere with each other:
\begin{enumerate}

\item P-even, $S_\chi$-independent
\ba
\CO_1 = \mathbf{1}, \ \ \ \ \ \CO_2 = (v^{\perp})^2, \ \ \ \ \ \ \CO_3 = i \vec{S}_N \cdot (\vec{q} \times \vec{v}^\perp) ,
\label{eq:ops1}
\ea
\item P-even, $S_\chi$-dependent
\ba
\CO_4 = \vec{S}_\chi \cdot \vec{S}_N, \ \ \ \ \ \CO_5 = i \vec{S}_\chi \cdot (\vec{q} \times \vec{v}^\perp) , \ \ \ \ \ \ \CO_6= (\vec{S}_\chi \cdot \vec{q}) (\vec{S}_N \cdot \vec{q}) ,
\ea
\item P-odd, $S_\chi$-independent
\ba
\CO_7 = \vec{S}_N \cdot \vec{v}^\perp,
\ea
\item P-odd, $S_\chi$-dependent
\ba
\CO_8 = \vec{S}_\chi \cdot \vec{v}^\perp, \ \ \ \ \ \ \CO_9 = i \vec{S}_\chi \cdot (\vec{S}_N \times \vec{q})
\label{eq:ops4}
\ea
\end{enumerate}
In addition, there are $T$-odd operators \cite{us}, but since these do not introduce any new nuclear responses we will not analyze their constraints separately.  

\subsection{Direct Detection Event Rates}

The calculation of direct detection rates in the general effective theory follows along essentially the same lines as in the case of standard spin-independent and spin-dependent interactions, 
which are reviewed in e.g. \cite{lewinsmith}.  However, we will see that there are a few non-trivial generalizations, and ultimately the final result will need to be parameterized differently from usual.  
For instance, it is common to parameterize direct detection rates in terms of the total wimp-proton cross-section, but more generally this quantity is momentum-dependent. The detector event
rate $R_D$ per unit time per unit energy per unit detector for a nucleus to recoil with energy $E_R$ is given by an average over the underlying dark matter velocity distribution
\be
\frac{dR_D}{d E_R}  &=& N_T \left\< n_\chi v_\chi \frac{d\sigma_T }{d E_R} \right\>,
\ee
where $n_\chi$ is the local DM density, $v_\chi$ is its velocity in the lab frame, $N_T$ is the number of nuclei per detector mass, and $\sigma_T$ is the scattering cross-section for dark matter and atomic nuclei. The kinematics
of scattering are easiest to follow in the center-of-mass frame, where the dark matter momentum $p$ and target momentum $k_T$ are equal and opposite, 
and given by
\be
&&\vec{p} = -\vec{k}_T = \mu_T \vec{v}_\chi , \qquad \qquad \textrm{(incoming)} \\
&& \vec{p}~' = - \vec{k}'_T = \vec{q} +\mu_T \vec{v}_\chi,  \qquad \textrm{(outgoing)}
\ee
where $\vec{q}=\vec{p}~'-\vec{p}$ is the momentum transfer, and $\mu_T$ is  the dark matter-nucleus reduced mass
\be
\mu_T &=& \frac{m_T m_\chi}{m_T+m_\chi}.
\ee
In this frame, energy conservation implies $|p|=|p'|$, so 
\be
\frac{q^2}{2} = p^2 - p \cdot p' = p^2 (1- \cos \theta) = \mu_T^2 v_\chi^2 (1- \cos \theta),
\ee
where $\cos\theta$ is given by the scattering angle in the center-of-mass frame.  Then, the measured recoil energy is just the kinetic energy transferred to the nucleus:
\be
E_R &=& \frac{q^2}{2 m_T} = \frac{\mu_T^2}{m_T} v_\chi^2 (1- \cos \theta).
\ee
The minimum velocity required for a given momentum transfer occurs at exact backscattering ($\cos \theta = -1$):
\be
v_{\rm min} &=& \frac{q}{2\mu_T}.
\ee
Thus, holding $E_R$ fixed and integrating the velocity over a halo distribution $f(\vec{v})$ implies the following formula for the event rate:
\be
 \frac{dR_D}{dE_R} &=& N_T \frac{\rho_\chi m_T}{m_\chi \mu_T^2 } \int_{\rm v_{\rm min}}  d^3 v  v^{-1}  f(\vec{v}) \frac{d \sigma_T}{d \cos \theta}.
\ee
The differential two-to-two cross-section is given as usual in terms of the matrix elements by averaging over initial spins and summing over final spins:
\be
\frac{d \sigma_T}{d \cos \theta} &=& \frac{1}{2j_\chi +1} \frac{1}{2j+1} \sum_{\rm spins} \frac{1}{32\pi} \frac{ |\CM|^2}{(m_\chi + m_T)^2} ,
\ee
where $j$ and $j_\chi$ are the spin of the nucleus and dark matter, respectively.  The calculation of these matrix elements depends on the 
ground state of the atomic nucleus, but the result can be
parameterized in terms of form factors for the various effective
theory operators.  We parameterize the resulting amplitude by 
\be
\frac{1}{2j_\chi+1}\frac{1}{2j+1} \sum_{\rm spins} | \CM|^2 &\equiv & \frac{m_T^2}{m_N^2} c_\CO^2 F_\CO(v^2, q^2).
\ee
The factor $m_T^2/m_N^2$ is explicitly factored out of the definition of $F_\CO$ for convenience, due to the difference in relativistic normalization
of states for nuclei $\< k_T|k_T'\> =(2\pi)^3(2m_T) \delta^{(3)}(k_T-k_T')$ vs. nucleons $\< k| k'\> = (2\pi)^3(2m_N) \delta^{(3)}(k-k')$.  More generally,
in the presence of multiple effective interactions, there may be interference terms, each with their own form factor:
\be
\frac{1}{2j_\chi+1}\frac{1}{2j+1} \sum_{\rm spins} | \CM|^2 &\equiv & \frac{m_T^2}{m_N^2} \sum_{ij} c_i c_j F_{ij}(v^2, q^2).
\ee
Thus, putting everything together, the generalized event rate $R_D$ per unit time  per recoil energy per detector mass is
\be
\frac{dR_D}{dE_R} &=& N_T \frac{\rho_\chi m_T}{32 \pi m^3_\chi  m_N^2 } \int_{\rm v_{\rm min}}  d^3 v  v^{-1}  f(\vec{v}) \sum_{ij} c_i c_j  F_{ij}(v^2, q^2).
\ee

\subsection{A worked example: Dark magnetic moment}

As a practical example of the use of the effective theory and the form factors described in the previous section,
we will now discuss in detail how to treat the case where dark matter couples to the Standard Model through
a dark magnetic moment.  That is, consider a dark sector that contains a massive gauge field $A_\mu'$ that
mixes kinetically with the photon
\be
\CL \supset \epsilon F'_{\mu\nu} F^{\mu\nu},
\ee
and furthermore, the leading interaction of the dark matter $\chi$ with $A_\mu'$ is through a magnetic moment
interaction.  The standard way of writing the magnetic moment interaction non-relativistically is as a coupling between the $\chi$ spin
$\vec{S}_\chi$ and the dark magnetic field $\vec{B}' = \vec{\nabla} \times \vec{A}$:
\be
\CL \supset 2 m_\chi \mu_{\rm DM} \frac{\vec{S}_\chi}{|S_\chi|} \cdot \vec{B}' .
\ee
This interaction by itself is not frame-independent, and must combine with other interactions to form a boost-invariant
combination.  When $\chi$ is spin-$\frac{1}{2}$, such a combination arises from the following Lorentz-invariant operator:
\be
\CL_{\rm int}=  \mu_{\rm DM} \bar{\chi} \sigma^{\mu\nu} \chi F'_{\mu\nu}.
\ee
 Integrating out the massive $A'_\mu$, one obtains the interaction between $\chi$ and
Standard Model matter:
\be
\CL \supset  \frac{\epsilon \mu_{\rm DM}e}{2m_A^2} ( \bar{\chi} i \sigma^{\mu\alpha} q_\alpha \chi ) j_{\rm EM}^\mu,
\label{eq:relMagMom}
\ee
where $j^\mu_{\rm EM}$ is the electromagnetic current. Restricted to protons and neutrons, it can be written
\be
j^\mu_{\rm EM} = \bar{p}(k') \left( \frac{(k+k')^\mu}{2 m_N} + \frac{g_p}{2} \frac{i\sigma^{\mu\nu}q_\nu}{2 m_N} \right) p(k) + \bar{n}(k') \left(  \frac{g_n}{2} \frac{i\sigma^{\mu\nu} q_\nu  }{2m_N} \right) n(k),
\ee
where $g_p=5.59$ and $g_n=-3.83$ are the proton and neutron magnetic $g$-factors, respectively.  
 Since coupling through kinetic mixing with the photon is a
particularly compelling class of models, and this  always leads to dark matter interactions through the EM current,
we discuss these more generally and in more detail in App.~\ref{sec:EMmoments}.  Continuing with the interaction
eq. (\ref{eq:relMagMom}), we can take its non-relativistic limit to obtain
\be
\CL_{\rm int} &=&   \frac{\epsilon \mu_{\rm DM}e}{m_A^2}\left[  \left( m_N q^2 \mathbf{1} + 4 m_N m_\chi i \vec{S}_\chi \cdot (\vec{q}\times \vec{v}^\perp) 
  + 2 g_p  m_\chi (q^2 \vec{S}_\chi \cdot \vec{S}_p - (\vec{q}\cdot \vec{S}_\chi)( q \cdot \vec{S}_p)) \right)_{\rm proton} \right. \nn\\
  && \left.+ \left( 2 g_n  m_\chi (q^2 \vec{S}_\chi \cdot \vec{S}_n - (\vec{q}\cdot \vec{S}_\chi)( q \cdot \vec{S}_n) )\right)_{\rm neutron} \right].
  \ee
The first line are interactions with the proton, and the second are with the neutron.  Comparing with the definitions of the effective theory operators
in Sec. \ref{sec:EFTlist}, we can read off the coefficients for this model:

\be
\textrm{protons}&:& c^{(p)}_1 = m_N q^2 G_M, \quad c^{(p)} _4 = 2 g_p m_\chi q^2 G_M, \quad  c^{(p)}_5 = 4 m_N m_\chi G_M, \quad c^{(p)}_6 = -2 g_p m_\chi G_M, \nn \\
\textrm{neutrons}&:& c^{(n)}_4 = 2 g_n m_\chi q^2 G_M, \quad   c^{(n)}_6 = -2 g_n m_\chi G_M, 
\ee
where
\be
G_M =   \frac{\epsilon \mu_{\rm DM}e}{m_A^2}.
\ee
Summing over the coefficient, and using the formulae in App. \ref{sec:reduction}, we find that
\be
\sum_{ij} c_i c_j F_{ij}(v^2, q^2) &=&  q^4 m_\chi^2 G_M \left[ m_N^2 \left( \frac{1}{m_\chi^2} + \frac{4 v^2}{q^2} - \frac{1}{\mu_T^2} \right) F_M^{(p,p)}(q^2) + 4 F_{\tilde{\Delta}}^{(p,p)}(q^2) \right. \\
&& \left. -2\left( g_n F_{\Sigma',\tilde{\Delta}}^{(n,p)}(q^2) +g_p F_{\Sigma',\tilde{\Delta}}^{(p,p)} (q^2)\right)
 + \frac{1}{4} \sum_{N,N'=n,p} g_N g_{N'} F_{\Sigma'}^{(N,N')}(q^2) \right] . \nn
 \ee

\section{Constraints and Gaps}
\label{sec:constraints}
\subsection{Qualitative discussion of Responses and Gaps}

We would like to first review the five responses of interest and provide a qualitative picture of the sensitivities of the various elements to each one.  Due to
interference, it will ultimately be useful to group the responses into sectors consisting of operators that can interfere with each other.  However,
let us first start by describing the novel responses individually.

\subsubsection{Standard spin-independent response (SI)}

The standard SI response, or $M_{p,n}$, is unchanged in our framework, and arises from $\CO_1$.  A convenient measure for discussing the strength of various operators
is just the interaction itself squared and averaged (summed) over initial (final) states, which we will label $(\CL_{\rm int}^2)_{\CO}$.  For $\CO_1$, this is 
\be
(\CL_{\rm int}^2)_{\CO_1} &\sim& K_N^2 ,
\ee
where $K_N$ is the coherence factor, defined as $(A-Z), Z$ for $N=n,p$.  

\subsubsection{Spin dependent responses (SD1 and SD2)}

 Even in the case of spin-dependent interactions, there are in fact two different ways that the nucleus can couple to spin.  The first one, SD1 (or $\Sigma''_{p,n}$), is a projection of spin
 in the direction parallel to the momentum transfer, $\vec{q}$, as in the case of operator $\CO_6$.  The second, SD2 (or $\Sigma'_{p,n}$), is a projection of the spin in a direction
 perpendicular to the momentum transfer, as for operator $\CO_9$.  Qualitatively, both these responses behave similarly to the standard SD response, favoring elements with unpaired
 nucleons in the outer shell ($p$ or $n$).  Thus,  $\CO_6$ and $\CO_9$ are roughly of the same relative size for different elements, except that they differ in their momentum-dependence:
 \be
(\CL_{\rm int}^2)_{\CO_6} &\sim& q^4 S_N^2 , \\
(\CL_{\rm int}^2)_{\CO_9} &\sim& q^2 S_N^2 . 
\ee
    However, quantitatively, SD1 and SD2 are different from each other and from the usual SD response (which arises from $\CO_4 = S_N \cdot S_\chi$ and is a linear combination of SD1 and SD2).  
In addition, as we will describe later, the SD2 response can interfere with responses sensitive to the angular momentum content of the nucleus (LD), producing important effects, which
can reduce the sensitivity of targets to DM in various regions of parameter space.  

\subsubsection{Angular-momentum dependent response (LD)}

Operators $\CO_5$ and $\CO_8$ contain the standard SI response $M_{p,n}$.  However, for nuclei which contain an unpaired nucleon with angular momentum, there can be an
important correction due to an angular-momentum-dependent response LD (or $\Delta_{p,n}$).   This correction is of order one for elements with unpaired protons,
such as $^{23}$Na and $^{127}$I, and for the isotopes with unpaired neutrons such as $^{73}$Ge and $^{131}$Xe. For $\CO_5$, the approximate interaction strength is
\be
(\CL_{\rm int}^2)_{\CO_5} &\sim& \frac{q^4}{m_N^2}  (L_N^2+ K_N^2 \frac{m_N^2}{\mu_T^2}).
\ee
Note that for $m_\chi \gtrsim m_T$, the ratio $\frac{m_N^2}{\mu_T^2}$ is approximately $1/A^2$, so that both terms are roughly comparable.  Also, of qualitative importance, is the fact that the
LD response can interfere with SD2,  reducing the sensitivity to DM in certain regions of parameter space.

\subsubsection{Angular momentum and spin dependent response (LSD)}

The dominant response for operator $\CO_3$ is one which is sensitive to a nuclear feature that has not been considered previously, namely the product of spin and angular momentum
$(\vec{L} \cdot \vec{S})_{p,n}$:
\be
(\CL_{\rm int}^2)_{\CO_3} &\sim & \frac{q^4}{m_N^2}  ((L_N \cdot S_N)^2 +S_N^2 \frac{m_N^2 }{ \mu_T^2}).
\ee
Thus we will refer to it as the LSD response (or $\Phi''_{p,n}$). As above, $\frac{m_N^2}{\mu_T^2}$ tends to suppress the second term, so the LSD response tends to dominate.
  To get a sense for this response, recall that when all $2(\ell+1)$ states of the spin-aligned ($j=\ell+\frac{1}{2}$) subshell
and all $2\ell$ states of the spin-anti-aligned ($j=\ell-\frac{1}{2}$) subshell are occupied, this dot product vanishes.  However, the $\ell= j + \frac{1}{2}$ orbital and $\ell=j-\frac{1}{2}$ have 
different energies, so one will start to fill before the other.  Taking $n_\pm(\ell)$ to be the occupation 
numbers of the $\ell \pm \frac{1}{2}$ orbitals, the expectation value $\< S_N \cdot L_N\>$ is proportional to $ (\ell+1) ~n_+(\ell)- \ell ~n_-(\ell) $.
Thus a mismatch between $n_\pm(\ell)$ of order $\ell$   
produces $\langle(\vec{L} \cdot \vec{S})\rangle \sim \ell_{highest}^2$ for most elements.   $\ell_{highest}^2$ grows with the atomic number of the nucleus, and
so this response tends to favor heavier elements.  Qualitatively, it is thus somewhat similar to the standard SI response, although important differences occur.  For example, 
$^{19}$F is far less responsive than $^{23}$Na, even though their SI properties are similar.  An additional important qualitative aspect of the LSD response is that it can
interfere readily with the standard SI response, should the latter have non trivial momentum dependence (as can be natural in UV relativistic models which generate $\CO_3$ in
the first place).

\subsection{Constraints on different responses}

\begin{figure}
\includegraphics[width=0.5\textwidth]{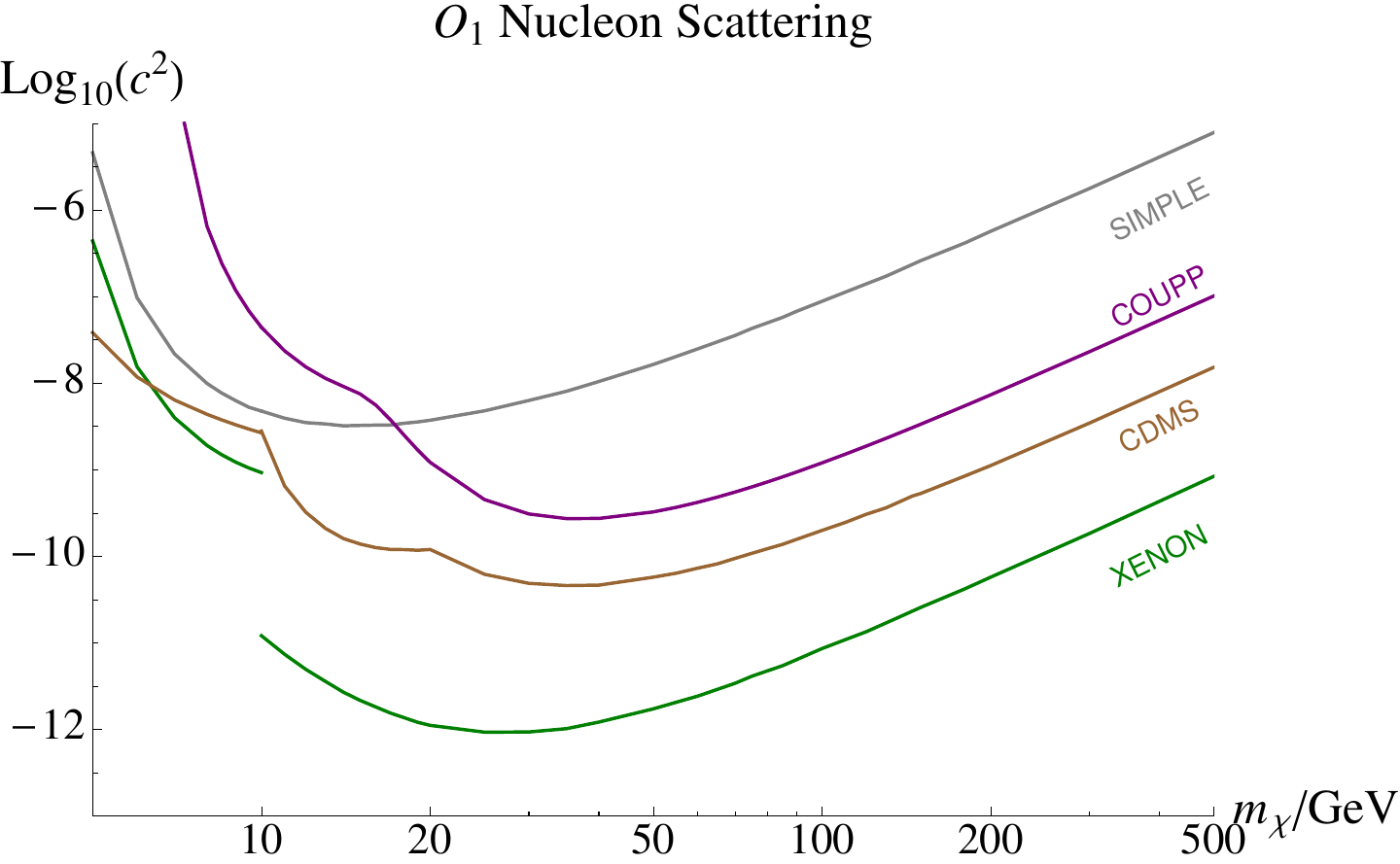}
\includegraphics[width=0.5\textwidth]{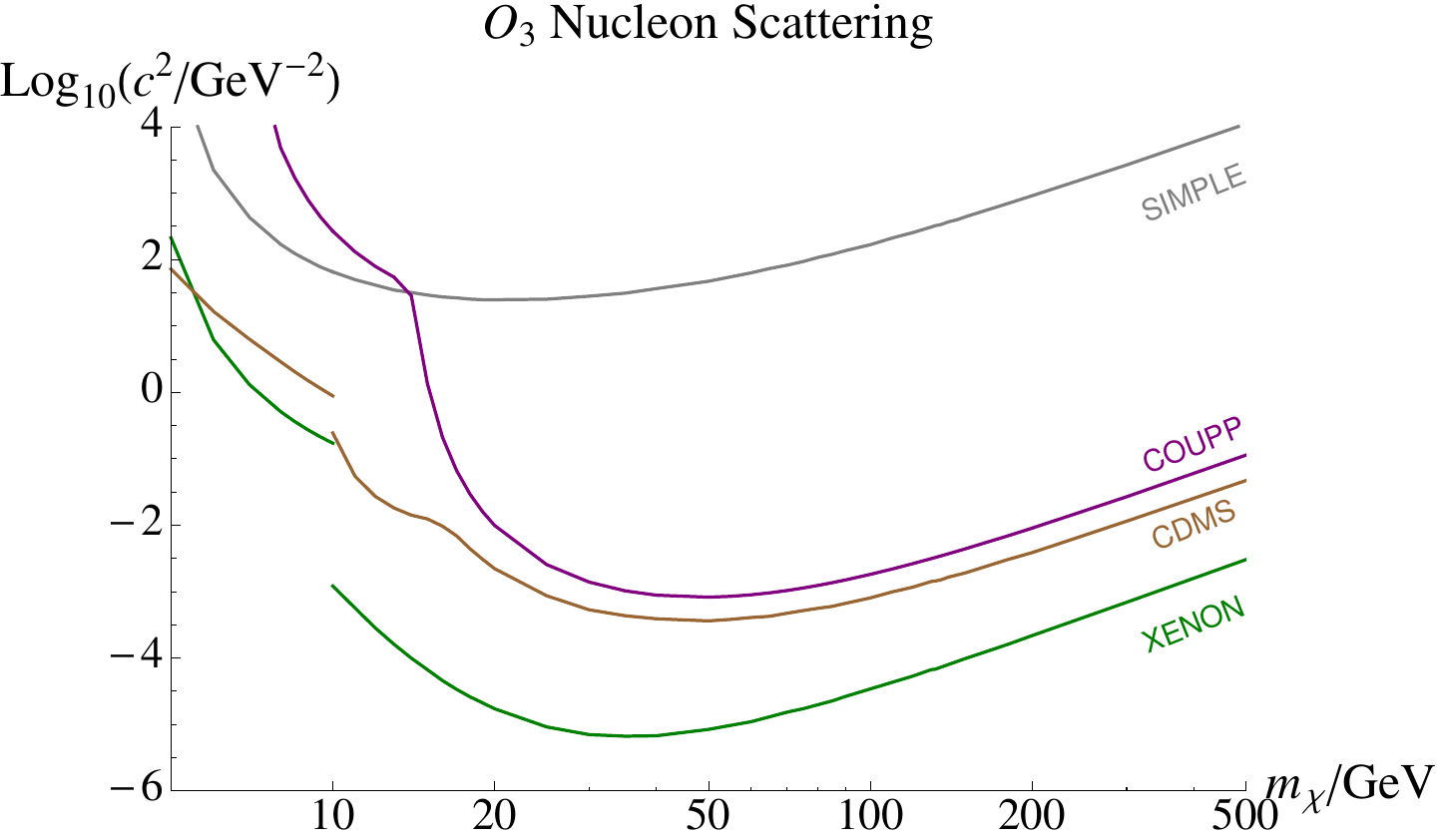}
\includegraphics[width=0.5\textwidth]{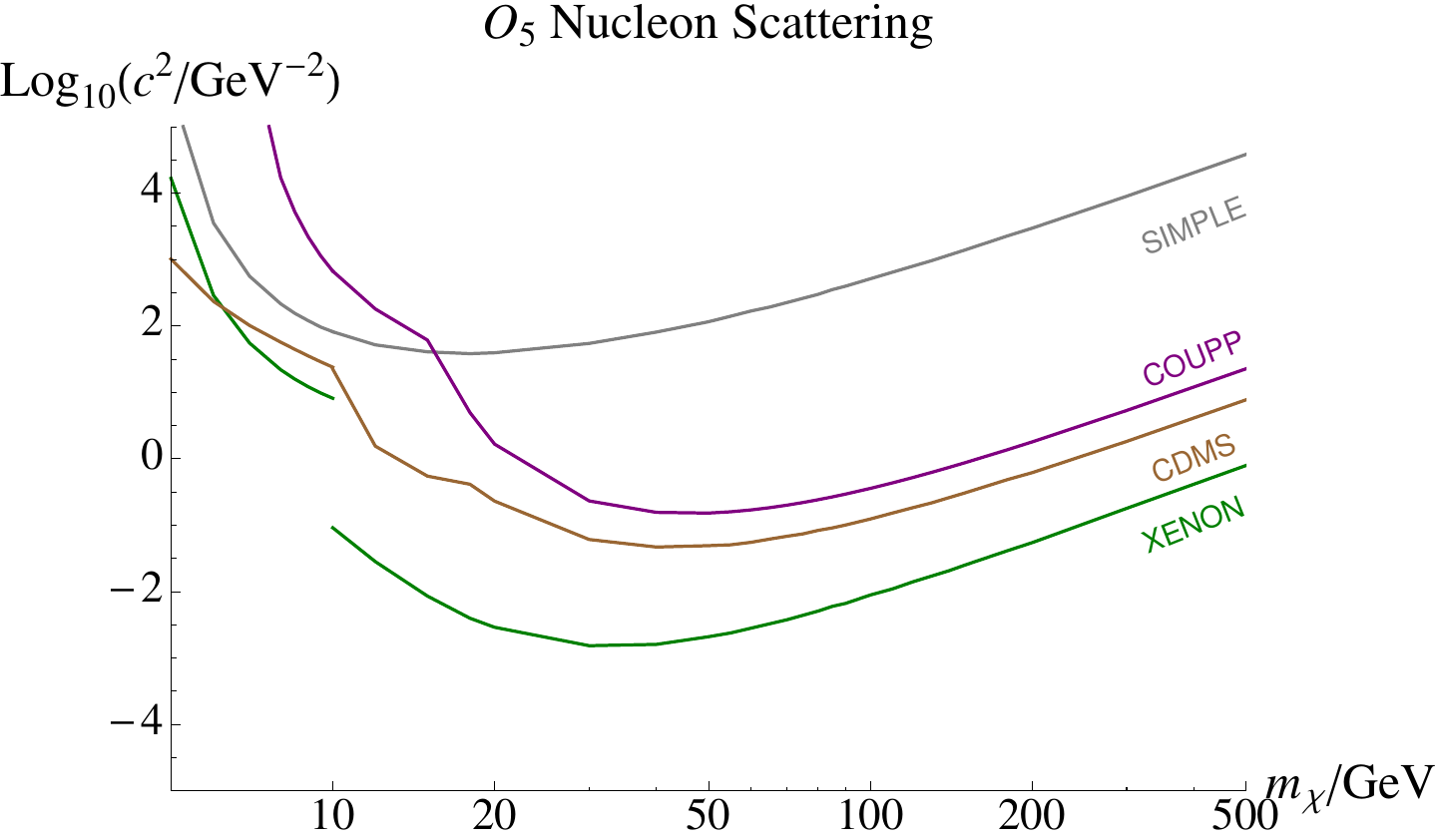}
\includegraphics[width=0.5\textwidth]{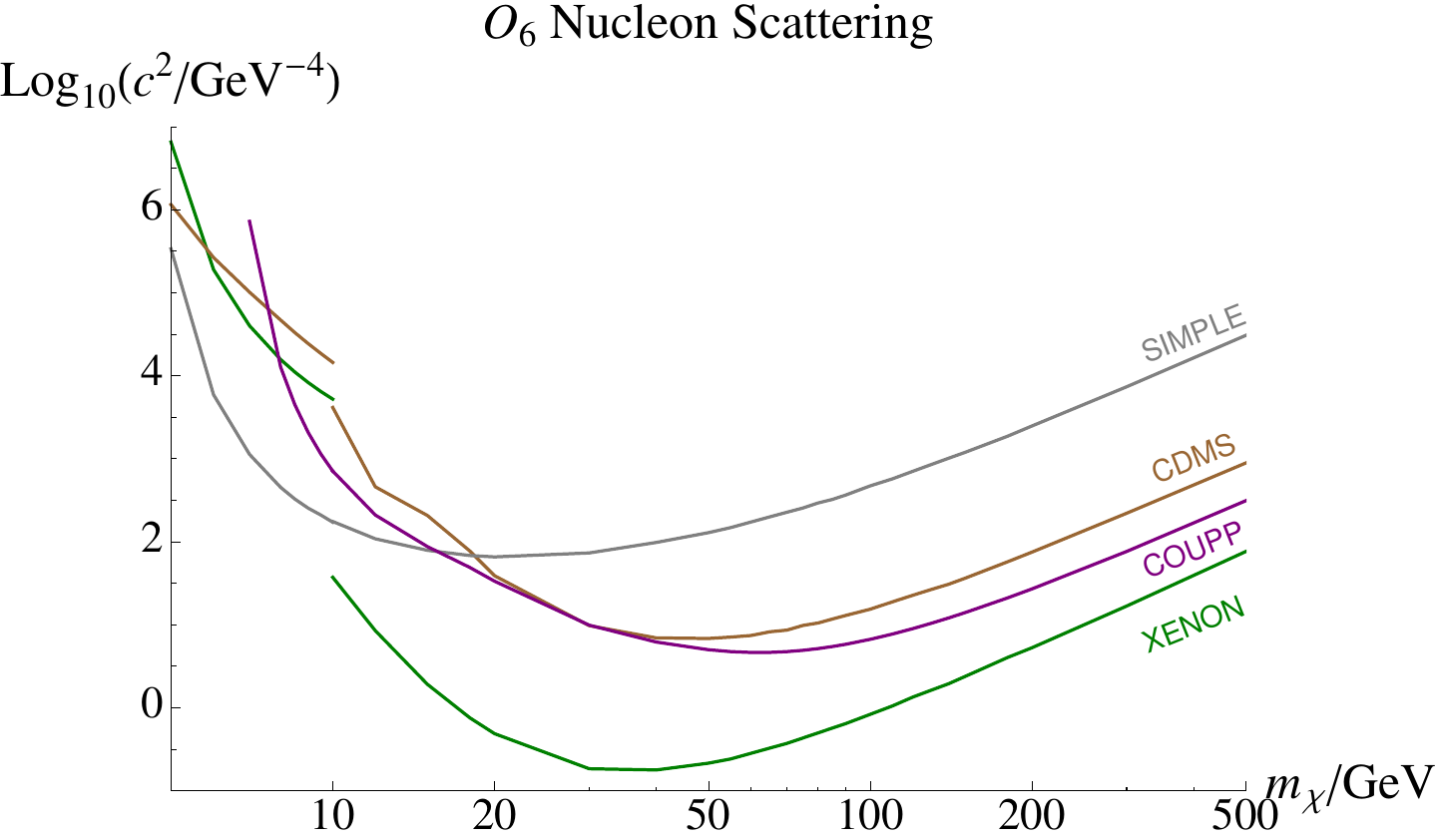}
\includegraphics[width=0.5\textwidth]{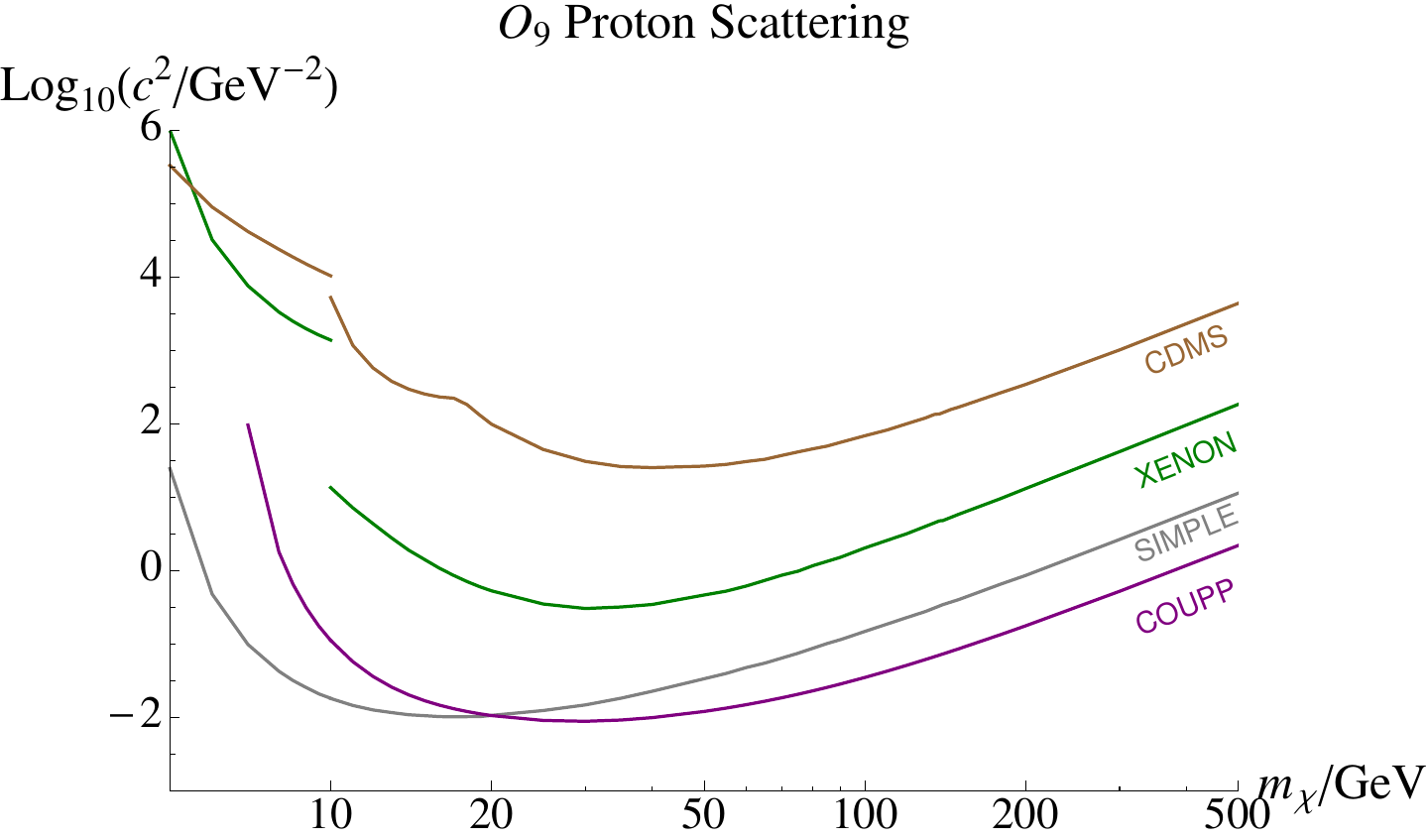}
\caption{Constraints on representative operators of the five independent nuclear responses, for each individual experiment. }
\label{fig:individualconstraints}
\end{figure}

Because the nuclear responses to the full set of effective theory operators in eqs.~(\ref{eq:ops1})-(\ref{eq:ops4}) depend only only a smaller set of 5 independent responses SI, SD1, SD2, LD, and LSD (i.e. $M,\Sigma'',\Sigma', \tilde{\Delta}$, and $\Phi''$), we will begin by showing the constraints on a representative effective theory operator for each of these.  These constraints will have overlap with those considered in \cite{fanEFT}, for which operators involving SI, SD1 and SD2 were considered. Fig. \ref{fig:individualconstraints}  therefore shows constraints from individual experiments on $\CO_1, \CO_3, \CO_5, \CO_6,$ and $\CO_9$.  By consulting eqs.~(\ref{eq:FFfirst})-(\ref{eq:FFlast}), one can read off that these five effective theory operators are sensitive to SI, \{LSD and SD2\}, \{SI and LD\}, SD1, and SD2, respectively. 
 Note that no operator is sensitive to LSD alone or LD alone - these always appear with SD2 or SI, respectively.  The following salient features emerge, in agreement with our above qualitative discussion of the different types of responses.  First, the strongest constraints in most cases is from XENON100, mainly due to its significant exposure and small background.  The only exceptions are $\CO_6$ and $\CO_9$ acting on protons, which are sensitive only to the relatively small proton spin in xenon-129 and xenon-131.  To separate out the effect of the large exposure vs. the intrinsic sensitivity of the isotopes to various operators, note the following experimental approximate exposures after including acceptances and efficiencies:

\begin{center}
\begin{tabular}{|c|c|}
\hline 
&  $\begin{array}{cc} \textrm{effective} \\ \textrm{exposure (kg d)} \end{array} $    \\
\hline
CDMS &  $\sim$ 200    \\
\hline
XENON100 &  $\sim$ 2500 \\
\hline
COUPP &  $\sim$ 25   \\
\hline
 DAMA & $\sim$ 50$/q$  \\
\hline
\end{tabular}
\end{center}
In the case of DAMA, the sensitivity is determined not by total exposure but by the amplitude of the irreducible modulating component that they observe, which depends on the quenching factor $q$.  Depending on whether the modulating component is due to scattering off of sodium or iodine, the appropriate quenching factor is $q_{\rm Na} \sim 0.3$ or $q_{\rm I} \sim 0.085$, respectively.  Similarly, COUPP sees 20 nuclear recoil events \cite{Behnke:2012ys}, which are all treated as potential dark matter events, so we choose their effective exposure in the above table to be their total exposure divided by 20. 
To obtain an estimate for the constraints, one can estimate the number of predicted events as
\be
\frac{dN}{d E_R} &\sim& 5000 {\rm keV}^{-1} \left( \frac{\rm exposure}{\rm kg \cdot day} \right) \left( \frac{ 100 \textrm{GeV}}{m_\chi} \right)^3 \CL_{\rm int}^2 ,
\ee
For the high-energy analyses, the the effective exposures above have all been defined so that the constraint is roughly $\frac{dN}{d E_R}  \lesssim 1 $.  

 In \cite{us}, various nuclear models were used to calculate the relevant nuclear form factors; we can read off from them the size of the interaction terms above:

\begin{center}
\begin{tabular}{|c|c|c|c|c|c|c|c|}
\hline 
& $S_n^2$ & $S_p^2$ & $L_n^2$ & $L_p^2$ & $(S_n\cdot L_n)^2$ & $(S_p \cdot L_p)^2$    \\
\hline
F & $8 \cdot 10^{-5}$ & 0.2 & 0.04 & 0.05 & 0.6 & 0.1  \\
\hline
Na & 0.0004 & 0.06  & 0.1 & 0.8 & 5.5 & 3.3\\
\hline
Ge & 0.02 & $ 5\cdot 10^{-6}$ &  1.1 & 0.003 & 35 & 100  \\
\hline
I & 0.004 & 0.07  & 0.4 & 2. & 100 & 500   \\
\hline
Xe & 0.02 & $2 \cdot 10^{-5}$  & 0.4 & 0.04 & 500 & 300  \\
\hline
\end{tabular}
\end{center}
All isotopes have been averaged over according to their natural abundance.  For comparison and to give a sense of the size of the uncertainties
involved in the calculation at small momentum transfer, the following table provides a summary of the quantities $S_p^2, S_n^2, L_p^2, L_n^2$
in various treatments in the literature.  Reference \cite{resselldean} present the results from two nuclear models, both presented below.  The differences
at fluorine and sodium are negligible, whereas the differences in the heavier elements - especially with protons for germanium and xenon and with neutrons for
iodine - can be quite significant.

\begin{center}
\begin{tabular}{|c|c|c|c|c|c|c|c|}
\hline 
& $S_n^2$ & $S_p^2$ & $L_n^2$ & $L_p^2$     \\
\hline
F \cite{vergados}&  8 $\cdot 10^{-5}$ & 0.2 & 0.04 & 0.05\\
\hline
Na \cite{resselldean} & 0.0004 & 0.06 & 0.1 & 0.08  \\
\hline
Ge \cite{ressell2} & 0.02 & 9$\cdot 10^{-6}$ & 1.0 & 0.02\\
\hline
I \cite{resselldean} $i$)&   0.006 & 0.09 & 0.6 & 2. \\
\phantom{I \cite{resselldean}} $ii$)&   0.004 & 0.1 & 0.4 & 2. \\
\hline
Xe \cite{resselldean} $i$)  & 0.04 & 0.0002 & 0.5& 0.02 \\
\phantom{Xe \cite{resselldean}}  $ii$) & 0.03 & 7 $\cdot 10^{-5}$ & 0.5& 0.05 \\
\hline
\end{tabular}
\end{center}

\subsection{Interference sectors}

We will now turn to exploring the full parameter space of interactions, where multiple interactions can contribute to scattering. As a result of interference between operators, when considering current experimental constraints, it is most useful to quote these constraints in terms of a bound on the 
sum of coefficients squared (i.e. $\sum_i c_i^2$) for each interfering sector.  In other words, we will be looking for flat directions within each sector, trying to determine
the size of the largest coefficients allowed in front of interfering operators of identical dimension.  We will restrict our analysis to the case where the operator coefficients in the effective theory are constants, independent of the momentum-transfer. In principle, one could add additional
$q^2$ dependence to the coefficients and parameterize the result as in
\cite{Feldstein:2009tr}. But we do not treat this case here.

One class of cancellations, coming from the interference in isospin-dependent interactions, has been considered in attempts to explain the current picture of direct detections (see e.g. \cite{Feng:2011vu}). In a model-independent framework, isospin is simply added to the parameter space of operators as a separate proton and neutron interaction coefficient, $c_p$ and $c_n$, for each operator. 
However, there is additional interference that occurs even between operators of the same isospin. In particular, the LD response interferes with the SD2 response (but both have the wrong parity to interfere with SD1).  The interference occurs entirely due to the fact that nucleon-DM 
operators containing the velocity operator $  \vec{v}^\perp$ are sensitive to the internal motion of nucleons within nuclei.  Roughly, one can think of this as due
to the fact that for the relative motion of a nucleon, $N$, around the nucleus, $ \vec{v}_{N,rel} \sim \frac{1}{m_N} \vec{q}\times\vec{L_N}$, where $\vec{L_N}$ is
the angular momentum of that nucleon. Additionally, the SI response interferes with the LSD response, because the expectation $\langle(\vec{L} \cdot \vec{S})\rangle$ is 
non-zero for most nuclei.

In practice, the constraints of course depend not only on the intrinsic sensitivity of the nucleus to the interaction but also on experimental setups such as exposures and efficiencies; a large exposure can compensate for a small intrinsic interaction strength. Calculating the constraints on the full parameter space will address the question of whether current experiments are complementary in probing the full range of interactions. 
We will also discuss in this section the importance of having a diversity of nuclear targets.
An important consideration in understanding constraints in the full parameter space is the dimensionality of interactions that an isotope can in principle be sensitive to.  This dimensionality is limited by the spin $j$ of the nucleus, since the maximum partial wave of a response that can contribute is $2j$. 
Thus, if this number is less than the dimension of the parameter space, we can always find a ``flat" direction along which the target is not sensitive due to interference. 
 In general this direction varies with $\vec{q}$ and will be ``smeared out" after we integrate over recoil energy.    For light elements like F, however, since $|\vec{q}|$ is small,  the event rate is populated only at low recoil energy, and thus the flat direction remains. 
This explains the fact that in the interference sectors (Fig. \ref{interferencesectors}) there is no constraint coming from SIMPLE.
In this case, another experiment that is sensitive in this direction will significantly improve the overall constraint. 

Similarly, we see the advantage of having many isotopes in the natural abundance of an
experiment. This is because the isotopes will differ in their sensitivity
to the various neutron interactions, with different sensitive
directions. Generically, the flat direction of one isotope will be
more sensitively probed by another, and hence the overall sensitivity for
the element will be better than the individual isotopes. Thus, the
experimental constraint will be stronger than the
constraint coming from each individual isotopic fraction. Furthermore, isotopes with higher spin constrain more directions in parameter space.  For example, out of the seven Xe isotopes, only $^{129}$Xe and $^{131}$Xe  have non-zero spin and can lead to constraints in the $\{\mathcal{O}_8, \mathcal{O}_9\}$ sector. The former of these,  $^{129}$Xe has spin 1/2, and consequently its form factors for SD1, SD2, and LD receive contributions only from their $J=1$ partial waves. E.g. for SD2 ($\Sigma'$)
\be
F_{\Sigma'}^{(N,N')} (q^2) &=& \frac{ 4\pi}{2j+1} \sum_{J=0}^{2j+1}  \< j || \Sigma_J^{\prime(N)} || j \> (q^2)\< j || \Sigma_J^{\prime (N')} | | j\>(q^2).
\ee
The $J=0$ partial wave vanishes for SD2, and since $j=1/2$, the sum truncates at $J=1$.  Consequently, viewed as a matrix $F_{9,9}^{(N,N')} = C(j_\chi) \frac{q^2}{16} F_{\Sigma'}^{(N,N')}$ manifestly has rank 1 and is thus sensitive only to one linear combination of $c_9^{(n)}$ and $c_9^{(p)}$. 
$^{131}$Xe, on the other hand, has spin 3/2 and its $F_{\Sigma'}^{(N,N')}$ is sensitive to two different directions. 

Our evaluations of the nuclear responses of common materials utilized as nuclear targets, F, Na, Ge, I and Xe, enable the calculations of expected event rates in this extended parameter space of non-relativistic operators. Here we describe in more details a scan of the four dimensional space parametrized by operator coefficients $c^{(N)}_i$  in the Lagrangian, for sectors $\{\mathcal{O}_1, \mathcal{O}_3\}$ and $\{\mathcal{O}_8, \mathcal{O}_9\}$, as well as for a single operator in each sector but allowing for isospin variations. 
Given that the targets vary in sensitivity along different directions in the parameter space, we will show the constraints imposed by various combinations of experiments XENON, CDMS,  SIMPLE and COUPP. That is, we will plot the intersection of the allowed regions of these experiments. For the first two experiments we consider separately high-energy and low-energy analyses \cite{CDMSlowER, Ahmed:2009zw, Sorensen:2010hv, Aprile:2012nq}.

\subsubsection{SI and LSD sector}

\begin{figure}
\includegraphics[width=\textwidth]{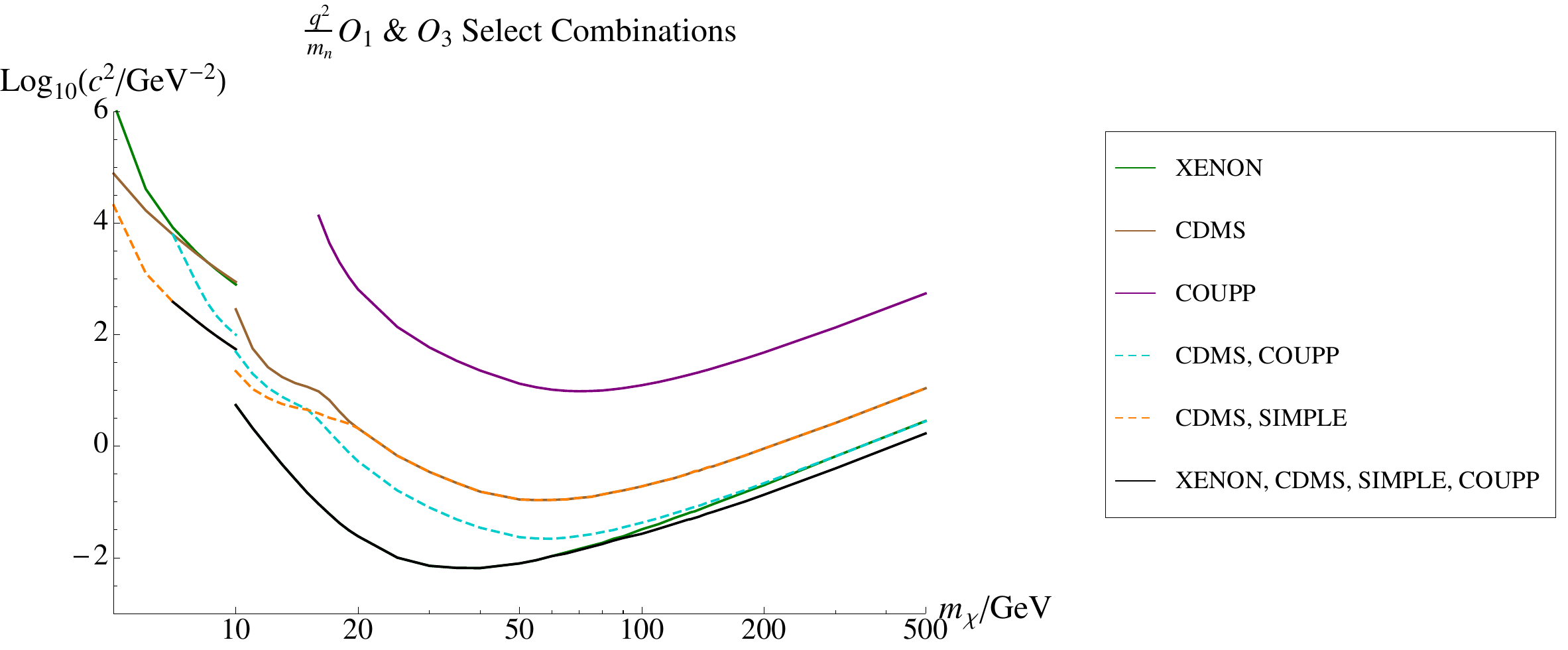}
\includegraphics[width=\textwidth]{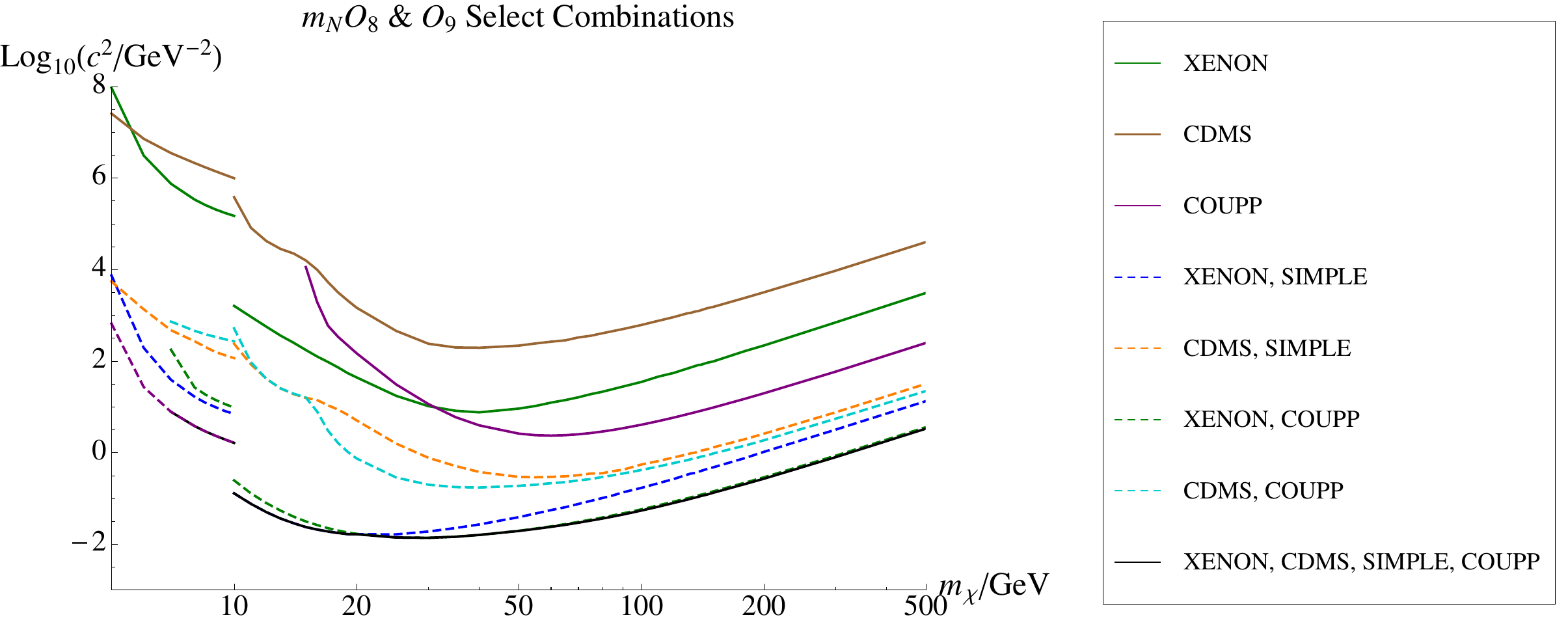}
\caption{Constraints on combinations of experiments for the two interference sectors.}\label{interferencesectors}
\end{figure}

Operators which are even under all symmetries, $\frac{q^2}{m_N} \CO_1$ and $\CO_3$ can interfere.  First, in Fig. \ref{interferencesectors}  we show the constraints on this sector (i.e. on $c^2 \equiv \sum_{N=p,n} c_{1N}^2 + c_{3N}^2$).
The various curves indicate which experiments are included to obtain a particular constraint.  
Because of its large exposure and sensitivity to SI and LSD responses, the constraint is being driven mostly by XENON100, and the combined limit does not change significantly when we subtract SIMPLE or COUPP.  At low $m_\chi <10$ GeV,  however, SIMPLE becomes constraining due to the  kinematics in DM-F nucleus collision. It pushes the constraint from XENON10 and CDMS (low threshold) significantly.  In order to gauge the importance of interference one can compare with the $\CO_3$ constraint in Fig. \ref{fig:individualconstraints}, where the constraint on $c^2$ is more severe, going down to $1/(100 \text{GeV})^2$ for certain DM masses. Thus, interference effects can lessen sensitivity by an order of magnitude or more
in the cross-section.

\subsubsection{SD2 and LD sector}

\begin{figure}
\includegraphics[width=0.8\textwidth]{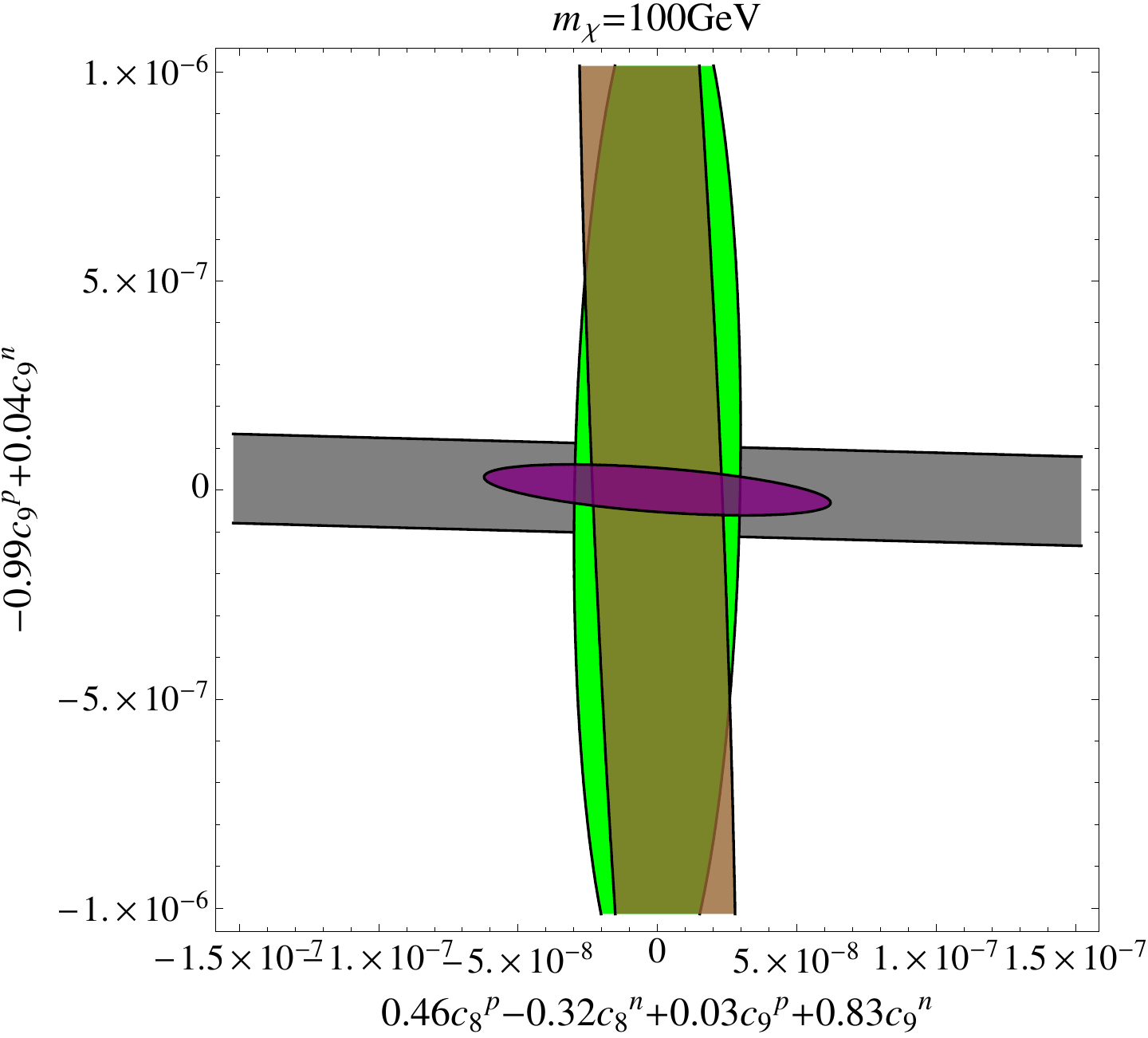}
\caption{A plot showing that for the $\{O_8, O_9\}$ sector, XENON (green) , CDMS (brown) and SIMPLE (gray), COUPP (purple) are sensitive in orthogonal directions.}\label{O89100GeV}
\end{figure}

Here there are two different sectors:  operators $q^2\CO_4$, $m_N\CO_5$, and $\CO_6$ are parity-even and comprise one sector, whereas $m_N\CO_8$ and $\CO_9$ are a similar, parity-odd sector. 
In each case, the operators containing LD (which is always accompanied by the standard SI coupling) are more heavily constrained than those containing SD interactions.  Thus, for example, $\CO_9$ is 
less constrained than $m_N\CO_8$, and thus interference between them does not significantly loosen the bound on $c^2$ (except for a small region near DM masses of 10GeV).  As is the case
with the standard SD case, experiments containing elements with unpaired neutrons (such as XENON100 and CDMS) need to be combined with those containing unpaired protons (such as SIMPLE and COUPP),
in order to guarantee greater sensitivity to DM. Indeed, we see an almost four orders of magnitude improvement in the constraint when SIMPLE or COUPP are combined with XENON or CDMS (Fig. \ref{O89100GeV}). 

Nevertheless, a comparison between single operator constraint (Fig. \ref{fig:individualconstraints}) and combined limits (Fig. \ref{interferencesectors}) shows that the combined limit comes mostly from the SD interaction.  The $F_M$ piece in $\CO_8$, albeit suppressed by DM velocity $v^2$, is still strong enough to impede $\CO_8$-$\CO_9$ cancellation. In this sector, interference has a minor effect.

\subsection{Flat directions and DAMA}

\begin{figure}
\includegraphics[width=0.5\textwidth]{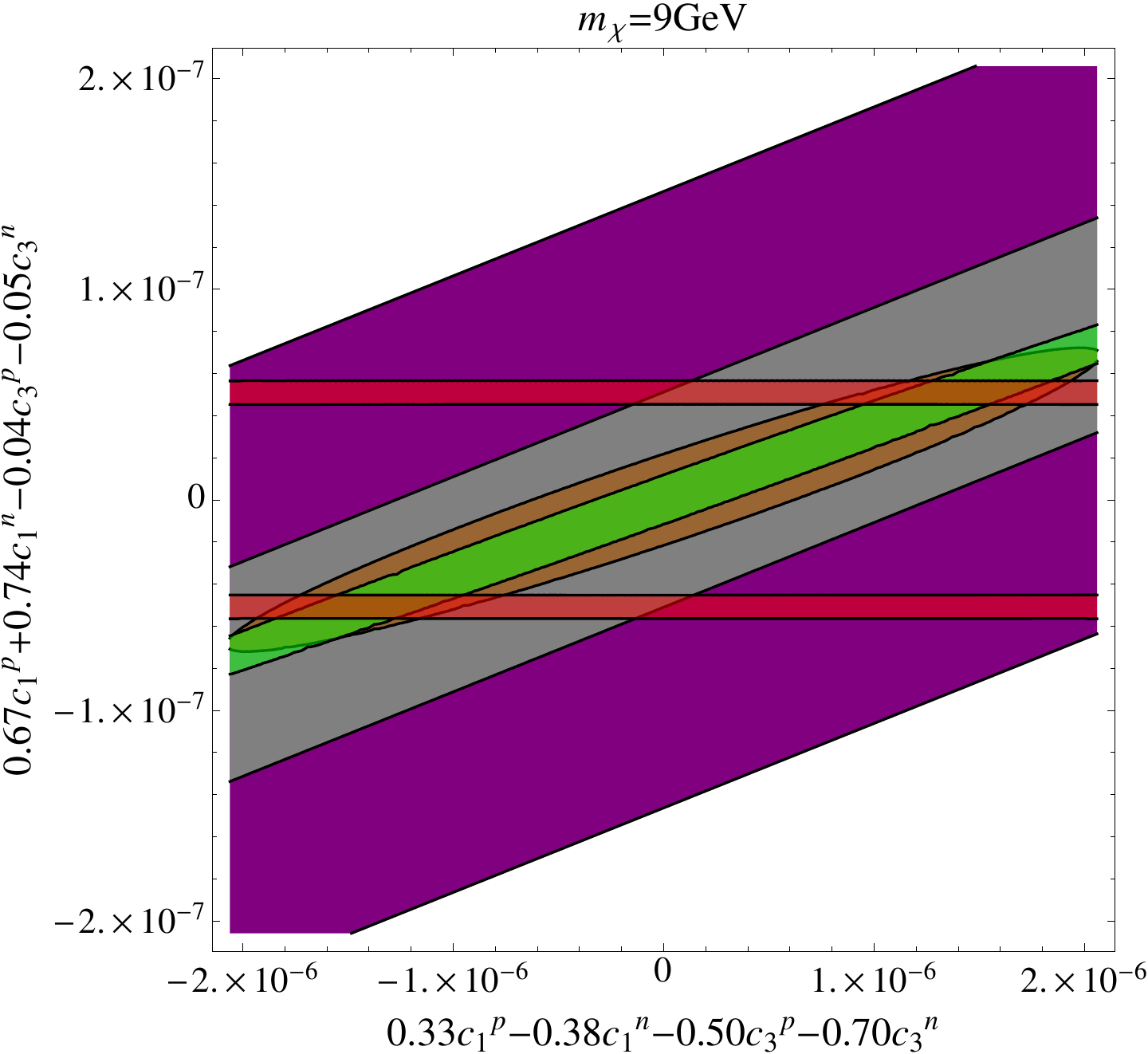}
\includegraphics[width=0.5\textwidth]{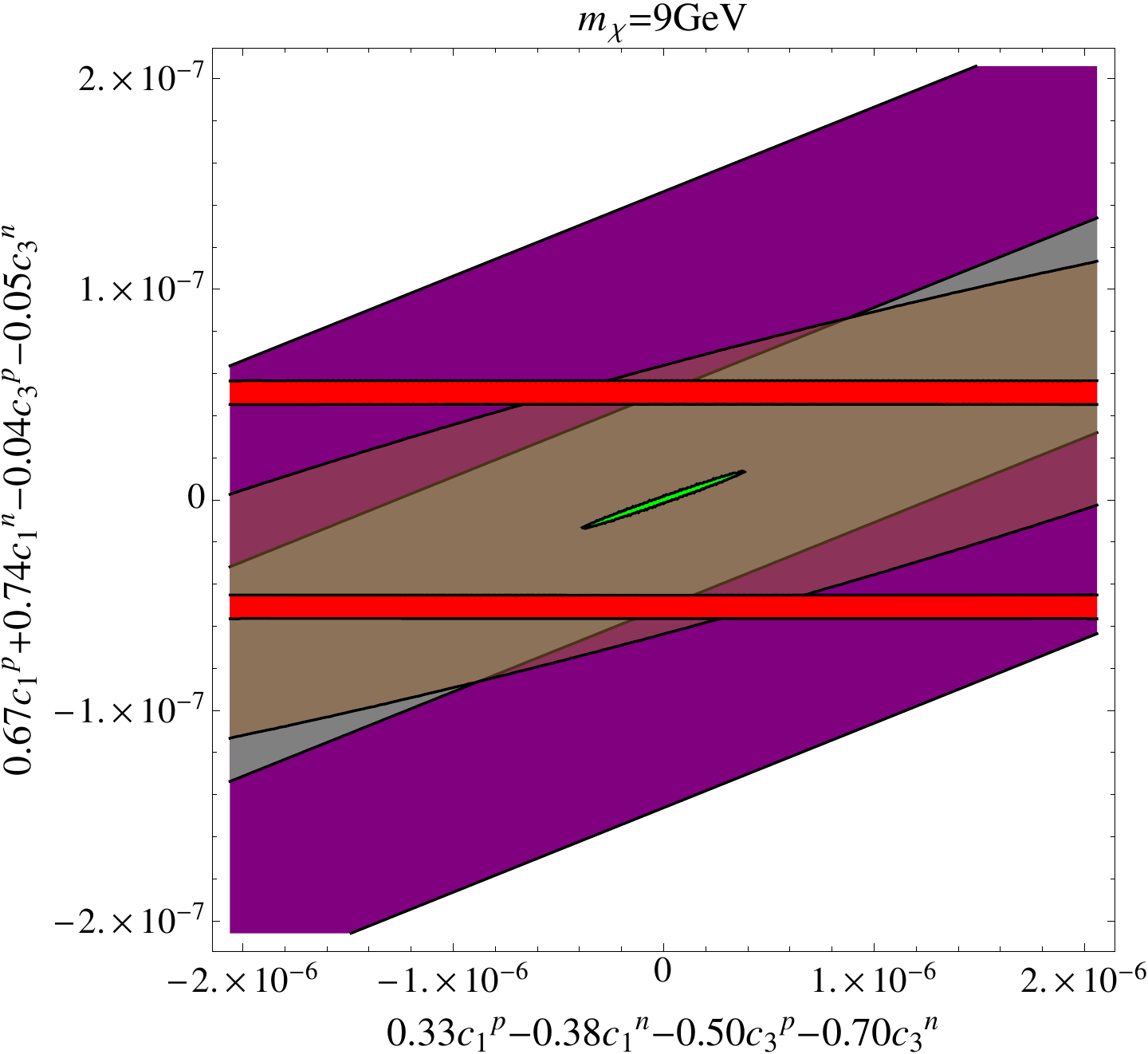}
\includegraphics[width=0.5\textwidth]{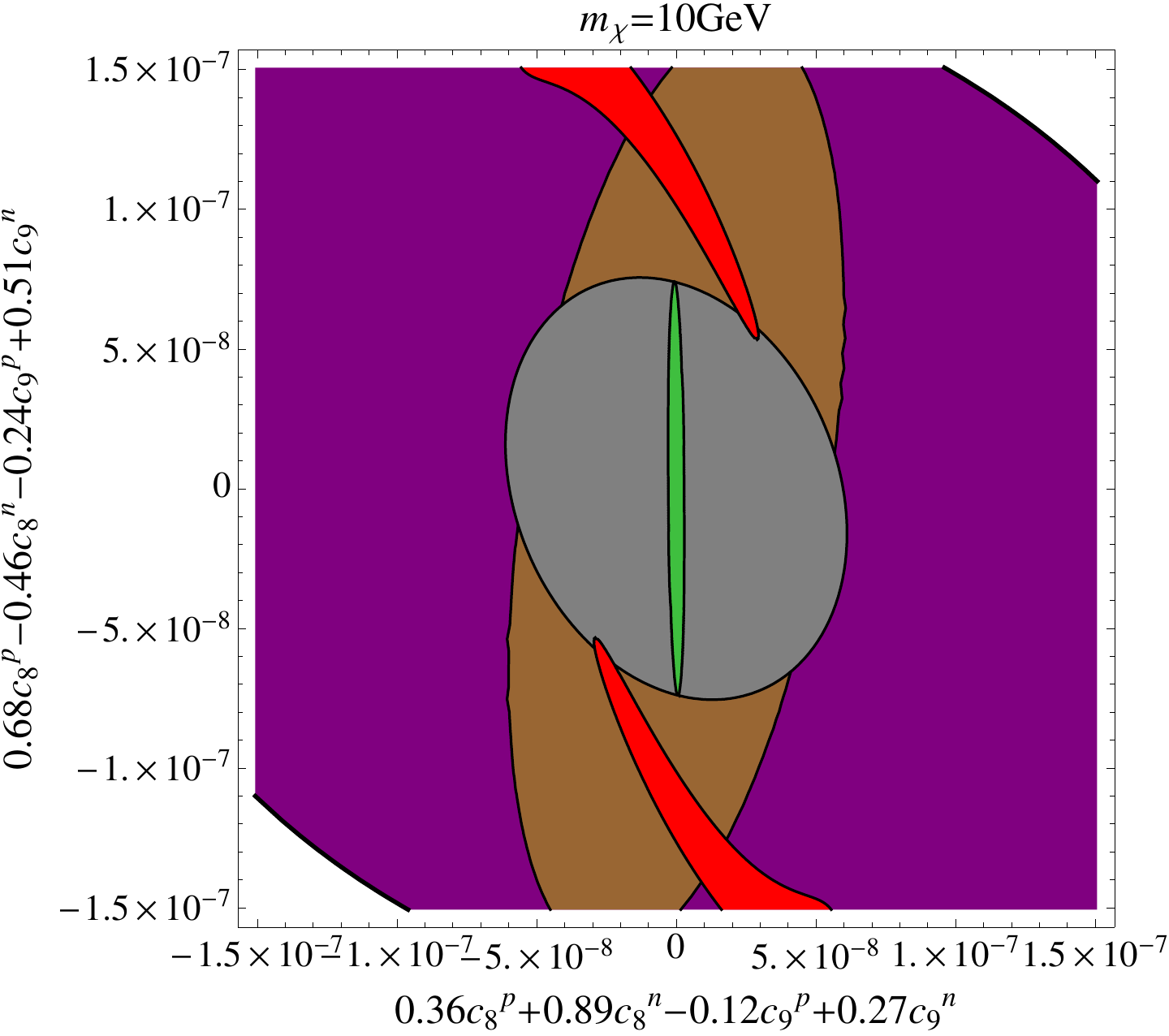}

\caption{
\label{fig:damacompatibility} The first two graphs are the low-energy experiment and high-energy experiment constraints on $\{O_1,O_3\}$ at $m_\chi=9$ GeV. The third graph is the $\{O_8,O_9\}$ sector at $m_\chi=10$ GeV, along which direction the DAMA  constraint region (red) is almost compatible with null experiments. Colors for the other experiments are as in Fig. \ref{O89100GeV}.}
\end{figure}

Several direct-detection nuclear recoil experiments have observed
positive signals that beyond their expected background
source.  While these signals could potentially disappear with better background modeling, it is nevertheless
interesting to try and interpret the anomalies as signals of WIMP scattering. To date, these experiments include DAMA \cite{DAMANew} , CRESST \cite{Petricca:2012ee}, and
CoGeNT \cite{COGENT2010}. 
Traditional analysis of spin-independent and
spin-dependent dark matter scattering yield a signal region which is
clearly incompatible with strong constraints coming from null
experiments. Using the effective field theory approach to dark matter
analysis, one can search through a larger parameter space to see if
there is a particular combination of operators which may resolve the
tension between DAMA and the null experiments. We have searched this
parameter space for regions which are compatible with the null
experiments and the annual modulation witnessed in the DAMA signal.

The DAMA detector uses NaI crystals, and the complementary kinematics of
the light and heavy atoms leads to multiple interpretations of the
signal from the detector, resulting either from sodium or iodine scattering. The
traditional spin-independent $O_1$ interaction thus yields two mass
ranges compatible with the DAMA modulation signal. 
   By including the full $\{O_1, O_3\}$ parameter space,
which includes nontrivial interference between operators, these mass ranges
are expanded to 8-13 GeV and 30-110 GeV. Similarly, for the $\{O_8,
O_9\}$ parameter space, the regions are from 8-17 GeV and from 25 GeV to and past the limit of our analysis, 500 GeV.
 These regions are only useful if they increase
the compatibility of DAMA with null experiments.

For the low mass ranges, the preferred region for DAMA is due to
sodium scattering. 
Sodium has nuclear spin $3/2$. Because each sector has a definite parity, only two of the orbitals can
contribute to scattering within a given operator sector (Either $\{O_1, O_3\}$
or $\{O_8, O_9\}$). Thus there are two ``flat directions'' and two
active directions in each sector. This allows for a large region of
compatibility with DAMA, and by exploiting this freedom it is
typically possible to fit DAMA with a single given null
experiment.

In the $\{O_1, O_3\}$ parameter space, the strong response of sodium
under the $\Phi^{''}$ form factor for low ( $<$ 10 GeV) dark matter
helps to reconcile DAMA with the null experiments. In the low-energy
analysis, there is a 1 GeV window centered on 9 GeV
which is compatible with CDMS, SIMPLE, COUPP, and XENON-10. (
  See Fig. \ref{fig:damacompatibility}.)  However, it is not possible to find a region which is
compatible with the very strong constraints coming from
XENON-100.

In the $\{O_8, O_9\}$ parameter space, it is again not possible to
find a region that is universally compatible. However, the strength of
the constraints in this parameter space is heavily dependent on the
complementary sensitivities of XENON and SIMPLE. If one does not
include either XENON or SIMPLE, it is possible to find regions which
are compatible with the three remaining null experiments. This
  is illustrated in Fig. \ref{fig:damacompatibility}.

\section{Discussion}
 \label{sec:discuss}

Although the results of direct detection searches are usually interpreted only in the standard spin-independent (SI) and spin-dependent (SD) scenarios, the full set of possible WIMP-nucleon interactions is much richer.  Recently, in \cite{us}, extending previous work by \cite{fanEFT}, we parameterized the full range of such interactions by constructing the non-relativistic effective field theory for dark matter direct detection. In this paper, we have considered analyses of the experimental constraints on this parameter space for elastically scattering WIMPs.  The major qualitative point of these interactions is that besides the SI and SD nuclear responses, there are angular-momentum (LD) and angular-momentum-and-spin (LSD) nuclear responses, which can differ significantly in their relative strengths at various nuclei from the SI and SD responses.  Furthermore, the SD response is a particular combination of two different possible spin-dependent responses, one (SD1) that couples to the longitudinal component of nucleon spin, and a second (SD2) that couples to the transverse component, and these have form factors that differ in detail.  

In section 2 of this paper, we have given a brief review of the effective theory itself and how event rates can be calculated for an arbitrary theory using the form factors in \cite{us} for the additional nuclear responses that appear.  It is our hope that this can serve as a useful guide to those wishing to perform an analysis themselves of the constraints on the effective theory or some subset of it.  In the remainder of the paper, we have analyzed the experimental limits on WIMP-nucleon cross-sections that obtain when one considers general points in the effective theory parameter space.  Our first set of results are the constraints on individual operators, shown in Fig. \ref{fig:individualconstraints}, where we have chosen one representative operator for each of the possible responses. We have also taken into account the importance of possible interference between different responses, which allows cancellations between different responses to occur that suppress the dark matter signal at particular isotopes.  Our results are presented as the constraints  on two different  subsectors of the effective theory.  The first subsector contains the SI, SD2, and LSD responses, while the second contains the SI, SD2, and LD responses; together, these contain all non-vanishing interference terms between different responses. These results are shown in Fig. \ref{interferencesectors}. One relevant question this enables us to address is to what extent the limits from combinations of experiments improve upon the limits of individual experiments themselves, or in other words, whether experiments can fill each other's ``gaps'' in sensitivity. We find that in the first subsector, the constraints from XENON100 are sufficiently strong that not much improvement is gained by combining them with additional experiments.  However, in the second subsector, we find that, mostly due to the smallness of the proton SD response in xenon, significant improvement is made by combining XENON100 with COUPP. 

 Finally, we have explored whether the anomaly at DAMA can be interpreted as dark matter signal consistent with all other experiments at any point in parameter space.  We find that, within our analysis assumptions, we rule out any possible elastically scattering solution.  However, certain highly tuned points in parameter space are ruled out only by a factor of 2 in cross-section, suggesting that with different analysis assumptions, in particular about the halo velocity distribution, a consistent interpretation might be allowed. 

\section*{Acknowledgments}
We would like to acknowledge useful conversations with Jared Kaplan, Michael Peskin, and Jay Wacker. ALF was partially supported by ERC grant BSMOXFORD no. 228169. WH is supported by the US Department of Energy under contract DE-SC00046548.  EK is supported by DOE grant DE-FG02-01ER-40676, NSF CAREER grant PHY-0645456, and also by an Alfred P. Sloan Fellowship.  NL and YX are supported by DOE grant DE-FG02-01ER-40676.
\appendix

\section{Experiments and analysis}

In this section we will briefly describe each direct detection
experiment included in our analysis. To perform a
conservative analysis we tag all reported events as WIMP
scatterings, except where background analysis is specifically mentioned. To calculate the allowed or 
  excluded regions, either $\chi^2$ statistics or maximum gap
methods \cite{Yellin:2002xd} are used based on the characteristics of the data
in each experiment.

\subsection{DAMA}
In our calculations, the DAMA signal is taken from 
\cite{DAMANew}. For sodium scattering events in NaI, we take a quenching factor of $Q_{Na}=0.3$, while for iodine, $Q_{I}=0.085$. No channeling effect is considered in this calculation. The total degree of freedom of the dataset we use is 17. The $90\%$ C.L. exclusion is therefore the contour with a $\chi^2 = 24.7$.

\subsection{CDMS II}
The CDMS II analysis used germanium detectors searching for WIMP interactions. In the data taken between July 2007 and September 2008, with an exposure of 612 kg$\cdot$days, two candidates events at recoil energies of 12.3 keV and 15.5 keV were observed \cite{CDMS2}. To calculate the 90\% C.L. exclusion, we use the pmax, or maximum gap method \cite{Yellin:2002xd}, namely setting the upper limit of the probability of seeing no events between the energies of the observed events and thresholds.

\subsection{CDMS low energy threshold analysis}
The CDMS collaboration reanalyzed their data taken between October 2006 and September 2008 to achieve a lower recoil energy threshold \cite{CDMSlowER}. The search took place within the energy region between 2 keV and 10 keV. Events in this region were previously rejected because of the high electron recoil background. In their analysis, the nuclear recoils were discriminated
from electron induced events based on the fact that electrons cause more ionization than nuclear recoils of the same energy. Events were identified as coming from a nuclear recoil based on their ionization energy and recoil energy, with cuts being determined by calibration with $^{252}$Cf sources.

We take the data as well as the nuclear-recoil acceptance efficiencies provided in the same analysis, and combine the events in the eight detectors together. These 400+ events are then divided exponentially with respect to the recoil energy into 20 bins. This method of binning is chosen as the measured event rate falls quickly with energy. A $\chi^2$ statistics is employed with a 90\% constraint at $\chi^2=28.4$. In order to account for the unknown background conservatively, only bins where the predicted event rate was greater than the observed rate were counted. Thus, in bins where the predicted rate is lower than the observation, the difference is posited to be due to background.

\subsection{XENON10 low energy threshold analysis}
The XENON10 collaboration also performed a low energy study of their
XENON10 data, down to the nuclear recoil energy of about 1 keVnr
(nuclear recoil energy) \cite{Sorensen:2010hv}. The experiment utilized
liquid xenon as the target, which, 
  containing an unpaired neutron, is more sensitive to neutron
induced scatterings. The detectors 
  recorded scintillation signal (S1) and ionization signal (S2). The
ratio of the two signals (S2/S1) acted as the criteria of background
rejection, since electron and nuclear events have different
values. However{\bf ,} the energy threshold of the scintillation efficiency $L_{eff}$ for XENON 10 was about 5 keV. To achieve lower threshold, they required only observations of the ionization signal and calibrated the events using the ionization yield $Q_{y}$, the ratio of S2 and recoil energy.

We extract the data from Figure 2 of their analysis, and the ionization
yield from Figure 1. Because of  systematic uncertainties, we
approximate conservatively, as suggested, $Q_{y}=4 \
\text{electrons/keVr}$ below $E_{nr}$ around 30 keVr, such that the
newly adopted $Q_{y}$ is a continuous function. Beyond the analysis
threshold there are five bins of data, 
  containing 41 events, among which, the four events with lowest
  $S2$ values are reported exactly. Since the $S2$ values for the events in the
    latter bins are unknown, we presume them distributed evenly in each bin so as to employ the $\chi^2$ method. 

The resolution of the experiment is taken into account as described in
\cite{Angle:2011th},  which presumes that
  the S2 signal follows a Poisson distribution. It is then
convolved with the expected event rate  from 1.4 keV, the cutoff in $Q_y$, to 60 keV, where the event rate becomes negligible.

\subsection{XENON100}
The XENON 100 data acquired for 13 months during 2011 and 2012, with 224.6 days $\times$ 34 kg exposure, exert by far the strongest constraint on the SI scattering cross section in the
WIMP parameter space \cite{Aprile:2012nq}. Two events at 7.1 keVnr  and 7.8 keVnr were detected, with background expectation of ($1.0 \pm 0.2)$ events.

As for the scintillation efficiency $L_{eff}$ , the parameter relating the nuclear recoil to electron equivalent energy and thus the photoelectrons, we directly use the data from the interpolation on measurements cited in this study. For nuclear recoil energies below 3 keVnr, there are no measurements of $L_{eff}$, and so we adopt the logarithmical extrapolation down to zero at 1 keVnr. Since the threshold energy in this experiment is 6.6 keVnr, the analysis will be relatively insensitive to the particular extrapolation used even after poisson fluctuations of the electron signal are taken into account.  Also, the resolution is $\sigma_{PMT} = 0.5$ photoelectrons \cite{Aprile:2011hx}, or translated to recoil energy, 1.4 keV. Thus the extrapolation of the scintillation efficiency under 3 keV has limited effect on the expected signal events. The uncertainty in $L_{eff}$ is neglected in our analysis.

The formulae of computing the expected events after considerations of statistics and resolutions is given in \cite{Aprile:2011hx}. We again impose the 90\% C.L. exclusion of our expected event rate on the XENON100 results with the maximum gap method.

\subsection{SIMPLE 2011}
The SIMPLE experiment applies superheated liquid C$_2$ClF$_5$ droplets
 in bubble chamber-like detectors, which is  particularly sensitive to spin-dependent
scattering due to the fluorine content. The first stage of the phase II SIMPLE dark matter search reported a 14.1 kg$\cdot$d measurement in 2010 \cite{Felizardo:2010mi}, where 14 events were observed. In 2011 the final results of the phase II experiment were presented \cite{Felizardo:2011uw}, which re-evaluated the data from the first
stage, identified 5 out of the previous 14 events as background, and
reduced the expected rate to 0.289 evts/kgd at 90\% C.L.. This imposed an even
stronger constraint on the cross sections than the newer result. A
refined efficiency is provided in the same study, which is included in
our calculations. We integrate over the expected event rate starting from the threshold energy of 8 keVnr to reach this value as the 90\% exclusion in the parameter space.

\subsection{COUPP 2012 results}

The COUPP experiment is also sensitive to spin-dependent scattering because of its CF$_3$I targets. The experiment reported results of the running from September 2010 to August 2011 with effective exposure of 437.4kg-days \cite{Behnke:2012ys}. Twenty single nuclear recoil event candidates were recorded. In our analysis, the efficiencies of the nuclear targets are taken from \cite{jhalltalk}. The observed events are presumed to follow the Poisson statistics, which are compared with the expected event numbers to set a $90\%$ exclusion limit.

\section{Electromagnetic moments in the effective theory}
\label{sec:EMmoments}

Dark matter interacting with the atomic nuclei through electromagnetic moments is a compelling and well-motivated special case of the general effective theory, and falls outside of the standard spin-independent and spin-dependent interactions.  Such electromagnetic moment interactions have been considered in the past (the earliest, to our knowledge, being \cite{pospelov}), but without the availability of the detailed nuclear response functions.  In this appendix, we will connect such interactions to our effective theory, which in turn allows one to calculate the corresponding momentum-dependent form factors. 

Restricted to the proton and neutron, the EM current is
\be
j^\mu_{\rm EM} = \frac{1}{2m_N} \left( K^\mu_p  + i \frac{g_p}{2} \sigma_p^{\mu\nu} q_\nu + i  \frac{g_n}{2} \sigma_n^{\mu\nu} q_\nu \right),
\ee
where $p$($n$) subscripts denote that the operator acts on protons(neutrons). Here, $g_p=5.59$ and $g_n=-3.83$. Taking the non-relativistic limit, the leading contributions to $j^\mu_{\rm EM}$ are
\be
j^0_{\rm EM} &=& 2 m_N \mathbf{1}_p , \\
j^i_{\rm EM} &=& K^i \mathbf{1}_p -i g_p \epsilon^{ijk} q^j S_p^k -i g_n \epsilon^{ijk} q^j S_n^k .
\ee
The first of these is (at this order) boost-invariant (this is just the statement that $j^0_{\rm EM}$ shifts by a term that is suppressed by powers of momenta), whereas the second is not. Under boosts shifting the velocity as $v^i \rightarrow v^i + \beta^i$, the current $j^i_{\rm EM}$ transforms according to
\be
j^i_{\rm EM} &\rightarrow& j^{i \prime}_{\rm EM} = j^i_{\rm EM} - \beta^i j^0_{\rm EM}.
\ee
To make a boost-invariant quantity, we can use $P^i$ or $K^i$, which transform according to
\be
P^i &\rightarrow& P^i - 2 m_\chi \beta^i , \qquad 
K^i \rightarrow K^i - 2 m_N \beta^i.
\ee
From this we can construct the following boost-invariant:
\be
\vec{J}_{\rm EM} &\equiv& \vec{j}_{\rm EM} - j^0_{\rm EM} \frac{\vec{P}}{2 m_\chi} 
=\left( - 2 m_N \vec{v}^\perp -i g_p \epsilon^{ijk} q^j S_p^k \right)_{\rm protons} + \left( -i g_n \epsilon^{ijk} q^j S_n^k\right)_{\rm neutrons} .
\ee
An obvious second choice besides $\vec{J}_{\rm EM}$ for a boost-invariant would be to use $K^i$ instead of  $P^i$ in the above definition.  However, this does not produce a new independent invariant, because it is related to invariants we have already constructed.  Specifically, 
\be
 \vec{j}_{\rm EM} - j^0_{\rm EM} \frac{\vec{K}}{2 m_N} = \vec{J}_{\rm EM} + j^0_{\rm EM} \vec{v}^\perp.
 \ee
 A similar discussion holds for the electric and magnetic fields $\vec{E}$ and $\vec{B}$ themselves. Since $q^0$ vanishes (at the order we consider), we in fact have
 \be
 \vec{E} &=& i \frac{\vec{q}}{q^2} j^0_{\rm EM}, \qquad
 \vec{B}= i \frac{1}{q^2} \vec{q} \times \vec{j}_{\rm EM},
 \ee
 so $\vec{E}$ is boost-invariant  but $\vec{B}$ is not:
 \be
 \vec{E} &\rightarrow& \vec{E}' = \vec{E} , \qquad
 \vec{B} \rightarrow  \vec{B}' = \vec{B} - \vec{\beta} \times \vec{E}.
 \ee
   This can be remedied as before by constructing a combination of $\vec{B}$ and $\vec{E}$ that is invariant:
   \be
   \vec{\mathcal{B}} &\equiv& \vec{B} - \frac{1}{2m_\chi} \vec{E} \times \vec{P} = i \frac{1}{q^2} \vec{q} \times \vec{J}_{\rm EM}.
   \ee
Now, it is straightforward to work out the form factor for a general electromagnetic moment interaction.  Consider the following electromagnetic moments:
\be
\CL &=& 2 m_\chi \left( \mu_{\rm DM} \vec{\mathcal{B}} \cdot \frac{\vec{S}_\chi}{|S_\chi|} + d_{\rm DM} \vec{E} \cdot \frac{\vec{S}_\chi}{|S_\chi|} + a_{\rm DM} \vec{J}_{\rm EM} \cdot \frac{\vec{S}_\chi}{|S_\chi|} + \frac{1}{6} e r_D^2 i \vec{q} \cdot \vec{E} \right),
\ee
which are (boost-invariant generalizations of) a magnetic dipole moment $\mu_{\rm DM}$, an electric dipole moment $d_{\rm DM}$, an anapole moment $a_{\rm DM}$, and a charge radius $r_D$, respectively.  Applying the above results, we see that the leading EFT coefficients for this Lagrangian are simply
\begin{enumerate}
\item Magnetic dipole moment:
\be
\CL &=&2m_\chi \mu_{\rm DM} \left(  \frac{g_p}{q^2 |S_\chi|} (\CO_6 - q^2 \CO_4) - \frac{2 m_N }{q^2 |S_\chi|} \CO_5 \right)_{\rm protons} \nn\\
&& + 2m_\chi \mu_{\rm DM} \left(  \frac{g_n }{q^2 |S_\chi|} (\CO_6 - q^2 \CO_4) \right)_{\rm neutrons} .
\ee

\item Electric dipole moment:
\be
\CL &=&2m_\chi d_{\rm DM} \left(  \frac{2m_N  }{q^2 |S_\chi|} \CO_{11} \right)_{\rm protons}  .
\ee

\item Anapole moment:
\be
\CL &=&2m_\chi a_{\rm DM} \left( -\frac{2 m_N }{|S_\chi|} \CO_8 + \frac{g_p }{|S_\chi|} \CO_9 \right)_{\rm protons} \nn\\
&& + 2m_\chi a_{\rm DM} \left( \frac{g_n }{|S_\chi|} \CO_9  \right)_{\rm neutrons} .
\ee

\item Charge radius:
\be
\CL &=&2m_\chi  e r_D^2 \left( -\frac{1 }{3}  m_N  \CO_1  \right)_{\rm protons} .
\ee

\end{enumerate}

\section{Form Factors for Effective Theory Operators}
\label{sec:reduction}
 The full set of form factors necessary
for the effective operators in eqs.~(\ref{eq:ops1})-(\ref{eq:ops4}) can be written in terms of a smaller number of basic independent ones as follows:
\begin{subequations}
\ba
F_{1,1}^{(N,N')} &=&  F_{M}^{(N,N')} , 
\label{eq:FFfirst}\\
F_{3,3}^{(N,N')} &=&  \left( \frac{q^4}{4 m_N^2} F_{\Phi''}^{(N,N')} + \frac{q^2}{2} \left( v^2 - \frac{q^2}{4 \mu_T^2} \right) F_{\Sigma'}^{(N,N')} \right), \\
F_{4,4}^{(N,N')} &=& C(j_\chi) \frac{1}{16}  \left( F_{\Sigma''}^{(N,N')} +  F_{\Sigma'}^{(N,N')} \right), \\
F_{5,5}^{(N,N')} &=& C(j_\chi) \frac{1}{4} \left( q^2 \left(v^2 - \frac{q^2}{4 \mu_T^2} \right) F_{M}^{(N,N')} +  \frac{q^4}{m_N^2} F_{\Delta}^{(N,N')} \right), \\
F_{6,6}^{(N,N')} &=& C(j_\chi) \frac{q^4}{16} F_{\Sigma''}^{(N,N')}, \\
F_{7,7}^{(N,N')} &=& \frac{1}{8} \left( v^2 - \frac{q^2}{4 \mu_T^2} \right)  F_{\Sigma'}^{(N,N')}, \\
F_{8,8}^{(N,N')} &=& C(j_\chi) \frac{1}{4} \left(  \left(v^2 - \frac{q^2}{4 \mu_T^2} \right) F_{M}^{(N,N')} +  \frac{q^2}{m_N^2} F_{\Delta}^{(N,N')} \right) ,\\
F_{9,9}^{(N,N')} &=& C(j_\chi) \frac{q^2}{16} F_{\Sigma'}^{(N,N')} ,\\
F_{1,3}^{(N,N')} &=& \frac{q^2}{2m_N} F_{M, \Phi''}^{(N,N')}, \\
F_{4,5}^{(N,N')} &=& -C(j_\chi) \frac{q^2}{8m_N}  F_{\Sigma', \Delta}^{(N,N')} ,\\
F_{4,6}^{(N,N')} &=&  C(j_\chi) \frac{q^2}{16}  F_{\Sigma''}^{(N,N')} ,\\
F_{8,9}^{(N,N')} &=& C(j_\chi) \frac{q^2}{8m_N}  F_{\Sigma', \Delta}^{(N,N')} ,
\label{eq:FFlast}
\ea
\end{subequations}
where 
\be
C(j_\chi) = \frac{4 j_\chi(j_\chi+1)}{3}.
\ee
All interference terms that are not listed above vanish.  $\CO_2$ is omitted as it does not appear at leading order from any relativistic interaction.

\bibliography{DMref}{}
\bibliographystyle{utphys}

\end{document}